\documentclass[fleqn,preprint,showpacs,preprintnumbers,amsmath,amssymb]{revtex4}
\usepackage{graphicx}
\usepackage{dcolumn}
\usepackage{bm}

\begin{document}

\title{Orbital Ferromagnetism and the Chandrasekhar Mass-Limit}
\author{M. Akbari-Moghanjoughi}
\affiliation{Azarbaijan University of
Tarbiat Moallem, Faculty of Sciences,
Department of Physics, 51745-406, Tabriz, Iran}

\date{\today}
\begin{abstract}
\textbf{In this paper we use quantum magnetohydrodynamic (MHD) as well as magnetohydrostatic (MHS) models for a zero-temperature Fermi-Dirac plasma to show the fundamental role of Landau orbital ferromagnetism (LOFER) on the magnetohydrostatic stability of compact stars. It is revealed that the generalized flux-conserved equation of state of form $B=\beta \rho^{2s/3}$ only with conditions $0\leq s\leq 1$ and $0\leq \beta< \sqrt{2\pi}$ can leads to a stable compact stellar configuration. The distinct critical value $\beta_{cr}=\sqrt{2\pi}$ is shown to affect the magnetohydrostatic stability of the LOFER ($s=1$) state and the magnetic field strength limit on the compact stellar configuration. Furthermore, the value of the parameter $\beta$ is remarked to fundamentally alter the Chandrasekhar mass-radius relation and the known mass-limit on white dwarfs when the star is in LOFER state. Current findings can help to understand the role of flux-frozen ferromagnetism and its fundamental role on hydrostatic stability of relativistically degenerate super-dense plasmas such as white dwarfs.}
\end{abstract}

\keywords{Landau orbital ferromagnetism, Chandrasekhar mass-limit, LOFFER, Spin-orbit Magnetization, Dipole force effect, Quantum magnetohydrodynamic, Magnetohydrostatic}
\pacs{52.30.Ex, 52.35.-g, 52.35.Fp, 52.35.Mw}
\maketitle

\section{Introduction}

\textbf{Since the pioneering discovery by Chandrasekhar in 1939 \cite{chandra1} concerning the mass-limit on compact stars and the hydrostatic stability mechanisms in such stars due to the relativistic degeneracy of electrons, there has been a growing interest towards the study of hydrodynamic properties of degenerate ionized matter, the so-called zero-temperature quantum plasmas \cite{bohm, pines, levine, Markowich, shukla}. It is well known that the matter under compression exerts enormous pressure called the degeneracy pressure due to the Pauli exclusion principle when the interparticle distances are lowered to become comparable to the de Broglie thermal-wavelength $\lambda_D = h/(2\pi m_e k_B T)^{1/2}$ \cite{landau}. Quantum peculiarities such as quantum tunneling, quantized Hall effect, magnetic quantization etc. ubiquitously appear as the degeneracy limit is reached. Such peculiar features prove to be of fundamental significance and applications in ordinary metallic and semiconductor materials. Many recent investigations based on the quantum hydrodynamics (QHD) and quantum magnetohydrodynamics (QMHD) models \cite{gardner, Marklund1, Marklund2, Brodin1, Brodin2, manfredi, haas1, haas2, akbari2, akbari3} indicate that the incorporation of the quantum electron-tunneling and degeneracy pressure can lead to quite different nonlinear dynamic effects in plasmas. It has also been remarked that the relativistic degeneracy caused by large-scale gravitational forces in stars which causes the gravitational collapse in stellar objects \cite{chandra2, chandra3} may also lead to distinctive nonlinear hydrodynamic features \cite{akbari4, akbari5, akbari7} due to the change in the thermodynamical quantities in the Fermi-Dirac statistics \cite{kothary}.}

\textbf{Among the greatest challenges today is the problems associated with the origin of strong magnetic fields present in many compact astrophysical entities such as white dwarfs, pulsars, neutron stars, etc. and its formidable role on the stellar chain of evolution. It is also believed that the magnetic field has a fundamental role in the formation and the dynamical processes in astrophysical environments \cite{weeler}. There has been extensive past studies on the thermodynamical behavior of degenerated electron gas under arbitrarily high magnetic field \cite{can1, can2, can3, can4, can5, can6, can7, can8}. Such investigations have revealed that the high magnetic field can lead to the anomalous quantization and spiky features in the electronic density of states (DoS) affecting all the thermodynamical properties of the Fermi-Dirac gas. It was suggested that under such a quantizing field the magnetic transverse collapse of the gas is possible \cite{chai, akbari1} where the Fermi-Dirac gas may become one-dimensional. A review of the current findings on the properties of matter in strong magnetic field has been reviewed in some recent literature \cite{dong, harding}. Recent studies based on the QMHD including the magnetization effect confirm the significant role of the electron spin-orbit magnetization effects on the nonlinear properties of degenerated quantum plasmas \cite{marklund3, marklund4, brodin0, brodin3, misra1, misra2, martin, mushtaq, zaman, vitaly, akbari9}. Particularly, a more recent study remarks distinctive paramagnetic nonlinear features of a Fermi-Dirac plasma due to the relativistic electron degeneracy \cite{akbari10}.}

\textbf{Another outstanding feature of a degenerated plasma under a strong magnetic field is that a ferromagnetic solution called the Landau orbital ferromagnetism (LOFER) is possible \cite{lee, con1, con2, con3, burk1, burk2} which may account for the large magnetic fields (as high as $10^8G$) estimated for some astrophysical compact objects. In the present investigation we use both magnetohydrodynamic and magnetohydrostatic (MHS) models to explore the role of Landau orbital ferromagnetism on compact stellar characteristics such as hydrodynamic quantum collapse, hydrostatic stability and the mass-radius relation comparing the results with that of previous ones. The presentation of the paper is as follows. The QMHD model including Bohm potential and the spin-orbit magnetization effects is introduced in Sec. \ref{equations} and the possible regimes for a quantum collapse is explored in Sec. \ref{calculation}. The hydrostatic stability of a LOFER state is investigated and the Chandrasekhar mass-limit is calculated based on the generalized LOFER equation of state in Sec. \ref{discussion}. Finally, a summary is given in Sec. \ref{conclusion}.}

\section{Magnetohydrodynamic Model and Degeneracy Collapse}\label{equations}

In this section we show that for a flux-conserved degenerate plasma the total pressure can vanish in the transverse direction making the MHD wave unstable due to the effect of plasma magnetization. In this case a transverse collapse may occur without the need for presence of the gravitational force. Let us consider the magnetohydrodynamic (MHD) equations for a completely degenerate dense \textbf{singly-ionized and quasineutral ($n_i\simeq n_e = n$) helium plasma with the center of mass density $\rho=m_i n_i + m_e n_e \simeq m_i n = 2m_p n$ ($m_p$ is the proton mass)}. We have for the continuity equation
\begin{equation}\label{cont}
\frac{{\partial \rho }}{{\partial t}} + \nabla \cdot(\rho {\bf{u}}) = 0,
\end{equation}
where, ${\bf{u}} = ({m_i}n_i{{\bf{u}}_{\bf{i}}} + {m_e}n_i{{\bf{u}}_{\bf{e}}})/\rho$ is the center of mass speed of plasma. Therefore, the generalized momentum equation including the quantum degeneracy pressure, dipole force, magnetization and electron nonlocality effects can be written in cgs units in the following form \cite{brodin0}
\begin{equation}\label{mom}
\rho \frac{{d{\bf{u}}}}{{dt}} = ({\bf{B}}\cdot\nabla) \left( {{\bf{B}} - {\bf{M}}} \right) - \nabla \left( {\frac{{{B^2}}}{2} - {\bf{M}}\cdot{\bf{B}}} \right) - \nabla {P_d} + \frac{{\rho {\hbar ^2}}}{{2{m_e}{m_i}}}\nabla \frac{{\Delta \sqrt \rho  }}{{\sqrt \rho  }},
\end{equation}
\textbf{in which $P_d$ is the electron degeneracy pressure and we have neglected the Bohm force on ions and the pressure due to them. The magnetization, $\bf{M}$ and the induced field, $\bf{B}$, are related through; $\bf{H}=\bf{B}-4\pi \bf{M(B)}$, with $\bf{H}$ being the magnetic field due to physical currents. The metastable Landau orbital ferromagnetism (LOFER) for the flux-conserved degenerate plasma model is given by $\bf{H}=0$ which leads to $\bf{B}=4\pi \bf{M(B)}$ \cite{sho}. On the other hand, it has been shown that the LOFER condition for a magnetized degenerate electron-gas leads to the field/density equation of state (EoS) of form; $B(r) = \alpha(\rho_6/\mu_e)^{2/3}$ \cite{con1}, where, $\rho_6=\rho/10^6$ and $\mu_e$ is the number of nucleon per electron (in this calculation we use $\mu_e=2$ for helium). The parameter $\alpha$ is the normalizing factor for magnetic field and will be found to be of the order $10^8$ for the white-dwarf mass-density ranges ($\rho_6\simeq 1$). However, this parameter is known to relate to some other parameters such as the plasma temperature, electron exchange interactions etc. \cite{burk1}. There has been many reports of compact star with strong internal or external magnetic fields \cite{crut, kemp, put, jor}. On the other hand, many reports confirm the role of flux-conservation on star formation \cite{bra} with the similar density dependence of magnetic field. In 1964 a theory based on flux conservation was suggested simultaneously by Ginzburg \cite{ginz} and Woltjer \cite{wolt} to explain the presence of intense magnetic fields in some young compact stars born in supernova explosions \cite{mestel}. As it will be apparent later in discussion the parameters $s$ and $\alpha$ in EoS of form $B = \alpha(\rho_6/\mu_e)^{2s/3}$ are central to the stability criteria of compact stellar configurations. There are some theoretical discrepancies on the values of these parameters. For instance, some calculations show that \cite{burk2} the value of $s$ in LOFER EoS, should be $4/3$ rather than $2/3$ calculated for this parameter previously \cite{con1}. Also, the calculations of the parameter $\alpha$ leads to different values due to the oscillatory nature of the spin-orbit magnetization elements. However, in this calculation we introduce a more general field/density EoS as $B = \alpha(\rho_6/\mu_e)^{2s/3}$ to show that only the restricted values of $0<s\leq1$ is consistent with the magnetohydrostatic (MHS) stability of compact stars. Also, we will find an upper limit on the value of $\alpha$ (or the strength of the magnetic field) for the known LOFER state ($s=1$) of stellar configuration.}

In a strongly magnetized Fermi-Dirac electron gas the equation of state (EoS) is quantized and we may write for the electron number-density in terms of Hurwitz zeta functions \cite{claud}
\begin{equation}\label{h}
\begin{array}{l}
n_e(x,\gamma ) = {n_c}{(2\gamma )^{3/2}}{H_{ - 1/2}}\left( {\frac{{{x^2}}}{{2\gamma }}} \right), \\
{P_{e\parallel}}(x,\gamma ) = \frac{{{n_c}{m_e}{c^2}}}{2}{(2\gamma )^{5/2}}\int_{0}^{\frac{{{x^2}}}{{2\gamma }}} {\frac{{{H_{ - 1/2}}(q)}}{{\sqrt {1 + 2\gamma q} }}} dq, \\
{H_z}(q) = h(z,\{ q\} ) - h(z, q + 1 ) - \frac{1}{2}{q^{ - z}}, \\
h(z,q) = \sum\limits_{n = 0}^\infty  {{{(n + q)}^{ - z}}} . \\
\end{array}
\end{equation}
where $h(z,\{q\})$ is the Hurwitz zeta-function of order $z$ with the fractional part of $q$ as argument and $P_{e\parallel}$ denote the degeneracy pressure parallel to the magnetic field. Note also that, $n_c=m_e^3 c^3/2\pi^2 \hbar^3$, $\gamma=B_0/B_c$ with $B_c=m_e^2c^3/e\hbar\simeq4.41\times 10^{13}G$ being the fractional critical-field parameter, $\varepsilon _{Fe}=\sqrt{1+x^2}=E_{Fe}/m_e c^2$ is the normalized Fermi-energy and $x=p_{Fe}/m_e c$ is the normalized Fermi-momentum the so-called relativity parameter. It is evident that, the electron degeneracy pressure is field dependent on both magnitude and direction so that $P_{e\perp}\neq P_{e\parallel}$. It has been shown that in such plasmas two distinct quantum and classical degeneracy regimes based on the parameter $x^2/2\gamma$ can be defined \cite{akbari11}. In our case the magnetic fields of interest are of the order $10^8G$ leading to the classical regime $x^2/2\gamma\gg 1$ for the white dwarf star density ranges. Hence, one may assuredly use the Chandrasekhar classical EoS for the degeneracy pressure as \cite{chandra1}
\begin{equation}\label{p}
{P_d(x)} = \frac{{\pi m_e^4{c^5}}}{{3{h^3}}}\left\{ {x\left( {2{x^2} - 3} \right)\sqrt {1 + {x^2}}  + 3\ln \left[ {x + \sqrt {1 + {x^2}} } \right]} \right\},
\end{equation}
\textbf{where, the well-known relativity parameter is $x=p_{Fe}/m_e c=(h/m_e c)(3n/8\pi)^{1/3}=(n/n_0)^{1/3}$ ($n_0=8\pi m_e^3 c^3/3h^3\simeq 5.9\times 10^{29}/cm^3$) \cite{kothary}, with $p_{Fe}$ being the relativistic Fermi momentum. We may also write the relativity parameter in terms of the plasma mass-density $\rho=2m_p n$ and $\rho_0=2m_p n_0=2m_p/3\pi^2\mathchar'26\mkern-10mu\lambda_c^{3}\simeq 2\times 10^6 gr/cm^3$ as $x=(n/n_0)^{1/3}=(\rho/\rho_0)^{1/3}$ with $\mathchar'26\mkern-10mu\lambda_c=\hbar/m_e c\simeq 3.863\times10^{-11}cm$ being the scaled electron Compton-wavelength. Note that in the forthcoming algebra we will use the normalized mass-density $\bar\rho=\rho/\rho_0$ dropping the bar notation for simplicity, hence, $x=\rho^{1/3}$.} Therefore, the normalized momentum equation, Eq. (\ref{mom}) may be cast in terms of the effective potentials \cite{akbari12} as
\begin{equation}\label{momq}
\rho \frac{{d{\bf{u}}}}{{dt}} = - \nabla \left( {{\Psi _m} + {\Psi _d}} \right) + \frac{{\rho {\hbar ^2}}}{{2{m_e}{m_i}}}\nabla \frac{{\Delta \sqrt \rho  }}{{\sqrt \rho  }},
\end{equation}
where $\Psi_m$ and $\Psi_d$ are the corresponding flux-conserved magnetic and electron degeneracy effective potentials (the normalization is given elsewhere \cite{akbari12}). Therefore, the inward magnetic-force, ${{\bf{F}}_m} = \nabla \Psi_m = - sC_s^2{\beta ^2}\nabla {x^{(4s - 3)}}/2\pi (4s - 3)$ (we have made use of $\alpha=\beta C_s$ with $C_s=c\sqrt{m_e/2m_p}$ being the quantum sound speed) and the outward degeneracy force, ${{\bf{F}}_d} = \nabla \Psi_d = {C_s^{2}}\nabla \sqrt {1 + x^2}$ oppose each other. It has been shown that on some conditions the total force is canceled giving rise to the quantum collapse for LOFER case \cite{akbari12}. However, for the general EoS considered here there exists a surface for which a quantum collapse can occur which is shown in Fig. 1 for the parameter range of $3/4\le s\leq1$. It is clearly observed that for a quantum degeneracy collapse to occur in the LOFER state ($s=1$) the value of $\beta$ should be above a critical value $\beta_{cr}=\sqrt{2\pi}$ \cite{akbari12}. It is also remarked that quantum degeneracy collapse is not possible for $s<3/4$. In the proceeding section we use the MHS stability of a magnetized degenerate plasma model to find some limits on the value of parameters $s$ and $\beta$ which is also a limit on the magnetic field of flux-conserved compact star.

\section{Hydrostatic Stability in Flux-Frozen Stellar Plasmas}\label{calculation}

Using a classical treatment, in this section, we consider the MHS equilibrium for homogenous spherical magnetized plasma. We also ignore the quantum tunneling effect on electrons due to large size of system compared to the interparticle distances. To begin with, let us consider a spherical shell-element of mass-density $\rho$ which is normalized to $\rho_0$ defined in previous section. The equilibrium for the shell-element is defined as $\bf{F}_{in}=\bf{F}_{out}$, where, the inward force consist of the gravity and the magnetic force and the outwards force is the electron degeneracy force. In a simpler argument we may write
\begin{equation}\label{eq1}
\frac{{d{P_{tot}}}}{{dr}} =  - \frac{{G\rho(r) M(r)}}{{{r^2}}},\hspace{3mm}{P_{tot}} = {P_d} + {P_m},
\end{equation}
where, $G$ is the gravitational constant and $M(r)$ is the mass of plasma within the radius $r$, defined as
\begin{equation}\label{eq2}
\frac{{dM(r)}}{{dr}} = 4\pi r^2\rho(r),
\end{equation}
where, again, $\rho$ is the normalized local density. Thus, the above definitions lead to the MHS stability condition of the form
\begin{equation}\label{eq}
\frac{1}{{4\pi {r^2}G}}\frac{d}{{dr}}\left( {\frac{{{r^2}}}{\rho(r) }\frac{{d{P_{tot}(\rho)}}}{{d\rho}}\frac{{d\rho(x)}}{{dx }}\frac{{dx(r) }}{{dr}}} \right) + \rho(r)  = 0,
\end{equation}
with the parameter $x=\rho^{1/3}$ being the Chandrasekhar relativity parameter. From the definitions for the degeneracy and magnetic pressures one obtains
\begin{equation}\label{pr}
\frac{{d{P_d}(\rho)}}{{d\rho }} = \frac{{C_s^2}}{6}\frac{{{x^2}}}{{\sqrt {1 + {x^2}} }},\hspace{3mm}\frac{{d{P_m}(\rho)}}{{d\rho }} = - \frac{s{C_s^2{\beta ^2}}}{{12\pi }}{x^{{{2s - 1}}}},
\end{equation}
where, we have used a general form of $B=\alpha\rho^{2s/3}$ ($s$ is real) for the magnetic equation of state previously defined. Hence, we have
\begin{equation}\label{prt}
\frac{{d{P_{tot}(\rho)}}}{{d\rho }} = \frac{{C_s^2}}{6}\left[ {\frac{{{x^2}}}{{\sqrt {1 + {x^2}} }} - \frac{{s{\beta ^2}}}{{2\pi }}{x^{{{2s - 1}}}}} \right].
\end{equation}
Thus, in terms of the relativity parameter, we rewrite Eq. (\ref{eq}) as
\begin{equation}\label{eqx}
\frac{{3\pi \mathchar'26\mkern-10mu\lambda _c^2{n_h^2}}}{{16{r^2}}}\frac{d}{{dr}}\left[ {{r^2}\left( {\frac{x(r)}{{\sqrt {1 + {x(r)^2}} }} - \frac{{s{\beta ^2}}}{{2\pi }}{x(r)^{{{2(s - 1)}}}}} \right)\frac{{dx(r)}}{{dr}}} \right] + {x(r)^3} = 0,
\end{equation}
where, $n_h=\sqrt{\hbar c/G}/2 m_p\simeq 1.3\times 10^{19}$ is the dimensionless hierarchy-number defined based on three fundamental constants, namely, the scaled Plank-constant, $\hbar$, the speed of light in vacuum, $c$ and the gravitational constant, $G$. It is clearly evident that, for the value of $s=1$ the magnetic field pressure gradient term in Eq. (\ref{eqx}) becomes independent of the relativity parameter.

Therefore, each stable model is obtained by integrating Eq. (\ref{eqx}) from the center ($r=0$) outwards until $x(r)$ vanishes for some $r=r_s$ which would be the surface (of the star). This is done by employing initial conditions $x(r=0)=x_c$ and $dx/dr(r=0)=0$ so that for every given value of the relativity parameter at the center ($x_c$) we find a distinct solution. However, it is convenient to put the Eq. (\ref{eqx}) in a more friendly shape by introducing dimensionless parameters $y=x/x_c$ and $\eta=r/r_c$ with $r_c=\sqrt{3\pi}n_h\mathchar'26\mkern-10mu\lambda_c/4x_c$ so that at the center ($\eta=0$) we will have $y=1$. Therefore, the Eq. (\ref{eqx}) in dimensionless form reads as
\begin{equation}\label{eqy}
\frac{1}{{{\eta ^2}}}\frac{d}{{d\eta }}\left[ {{\eta^2}\left( {\frac{{{x_c}y(\eta)}}{{\sqrt {1 + x_c^2{y(\eta)^2}} }} - \frac{{s{\beta ^2}}}{{2\pi }}x_c^{2(s - 1)}{y(\eta)^{2(s - 1)}}} \right)\frac{{dy(\eta)}}{{d\eta }}} \right] + {y(\eta)^3} = 0,
\end{equation}
It is noticed that at the very high central density ($x_c\rightarrow \infty$) Eq. (\ref{eqy}) reduces to the famous Lane-Emden equation of index 3, which for $\beta=0$ (unmagnetized case) leads to the well-known Chandrasekhar mass-limit of $M_{Ch}\simeq1.43M_S$ ($M_S$ being the mass of the sun). A standard integration algorithm such as the Runge-Kutta fourth-order (as I have carried out the calculations with mathematica software) may be used to find, for instance, the value of $\eta=\eta_s$ at which $y$ vanishes (the surface of star) from which the radius and the mass of the stable configuration (star) can be calculated via the following relations
\begin{equation}\label{rm}
{r_s} = r_c{\eta _s},\hspace{3mm}M = 4\pi \int_0^{{r_s}} {\rho(r) {r^2}dr}  = \frac{{\sqrt {3\pi } }}{4}{m_p}n_h^3\int_0^{{\eta _s}} {{y(\eta)^3}{\eta ^2}d\eta}.
\end{equation}
It is interesting, however, to note the unique scaling, i.e. $n_h\mathchar'26\mkern-10mu\lambda_c\simeq 0.787 R_E$ ($R_E$ being the earth's radius) and $2m_p n_h^3\simeq 1.849 M_S$ ($M_S$ being the sun's mass). These relations for $\beta=0$ (unmagnetized) case, following the Chandrasekhar's pioneering work, has been reviewed in Ref. \cite{gar}. The main goal of next section is to explore the effect of Landau orbital ferromagnetism on magnetohydrostatic stability and consequences on the Chandrasekhar mass-limit in completely degenerate magnetized plasmas.

\section{The Chandrasekhar Limit for Flux-Frozen Compact Star}\label{discussion}

Equation (\ref{eqx}) can be solved for given values of $\beta$, $s$ and $x_c$. The solution $y(\eta)$ for the simplest case of $\beta=0$ (or $s=0$) which corresponds to the unmagnetized (uniformly magnetized) plasma case is shown in Fig. 2 for different values of central fractional density $x_c$. The values of $\eta_0$ where $y(\eta_0)=0$ indicate the surface of plasma from which the radius and mass of the configuration can be calculated. It is observed that as the value of $x_c$ is increased $\eta_0$ also increases and in the limit it approaches a liming value $\eta_{Ch}$ called the Chandrasekhar value. Figure 3 shows the famous Chandrasekhar mass-radius plot (in terms of sun's mass and earth's radius) and the corresponding Chandrasekhar mass-limit $M_{Ch}\simeq1.43M_S$ for unmagnetized case. Following is the evaluation of the problem for different values of parameters $s$ and $\beta$.

\subsection{Case $s>1$}

For the case $s>1$, it is observed (e.g. see Fig. 4 for $s=2$) that, as one increases the value of $x_c$ to $x_c=1$, the value of $\eta_0$ increases and approaches to a liming value. However, it is remarked that, there are no stable solutions for $x_c>1$ in this case. In other words, in this model, i.e. $s>1$, no dense-centered stable configuration such as observed for a star is allowed. Therefore, this case can not correspond to a physical solution.

\subsection{Case $s<0$}

On the other hand, for negative values of $s$, when $x_c$ is finite no $\eta_0$-value can be obtained, hence, the plasma must be infinite. However, for $x_c=+\infty$ the same Chandrasekhar-limit is obtained, regardless of the value of parameter $\beta$, as it is evident from Eq. (\ref{eqx}). For instance, Fig. 5 shows the solution $y(\eta)$ and the variation with respect to different values of $x_c$ for the case $s=-1$, indicating that, the case $s<0$ is unphysical.

\subsection{Case $0<s<1$}

For the range of $0<s<1$ the plasma is stable for large values of $x_c$. However, as $x_c\rightharpoonup 0$ it is observed from Eq. (\ref{eqx}) that the integrand diverges and the plasma becomes unstable. Therefore, in this model there may be a radius-limit to the plasma evaluation of which is beyond the scope of this paper. Figure 6 for $s=1/2$ shows the similar features as Fig. 2 as the central density of plasma is increased. Also, evaluation of Eq. (\ref{eqx}) reveals that the same Chandrasekhar mass-limit is obtained and that in this model the mass-limit is invariant under the change in $\beta$ parameter (e.g. see Fig. 7). Another important feature in this model is the existence of critical value for $\beta$. Evaluation of the solutions to Eq. (\ref{eqx}) for this case shows that stable configurations exist only below a critical value $\beta_{cr}=2\pi$.

\subsection{The LOFER Case $s=1$}

In this model which corresponds to the equation of state of form $B=\beta\rho^{2/3}$ and is known as LOFER state, unlike the previous case, the plasma is stable for the whole range of the parameter $x_c$. However, the stability of the plasma is removed for values of $\beta>\beta_{cr}$. It is also remarked from Fig. 8 that in this model the chandrasekhar-limit decreases as the value of $\beta$ is increased. From the previous discussion it may be concluded that the only stable and physical model corresponds to the cases $s=1$ and $\beta<2\pi$. The magnetogravity collapse condition ($\beta<2\pi$) is to be compared with the one obtained for the quantum collapse mentioned above ($\beta>2\pi$). Therefore, in a dense ferromagnetic plasma depending on the value of $\beta$ only one of possible collapses is possible of which the magnetogravity collapse can occur in hydrostatically stable configurations. It is noted that, although the LOFER-state is a metastable configuration and requires a sufficient condition to operate, however, once it occurs the quantum collapse will inevitable. The variation of Chandrasekhar limit with the LOFER parameter $\beta$, shown in Fig. 9 confirms that the mass-limit decreases with the increase in the value of $\beta$ until it vanishes at critical value $\beta=\sqrt{2\pi}$. Also, Fig. 10 compares the Chandrasekhr mass-radius curves for the tentative values of $\beta^2/2\pi=0,0.1,0.5$ and $s=1$.

It is remarked that, the flux frozen assumption used in this investigation can lead to fundamental effects of the Chandrasekhar mass-radius relation for the case of $s=1$ and the mass-limit on magnetic white dwarfs can be much lower than $1.43217$. This stellar model is also consistent with supernovae type I theory of Hoyle and Fowler \cite{hoyle} which requires stellar mass between 1.01 and 1.40 solar mass. It is obvious that for a complete treatment of stellar stability the effects such as Thomas-Fermi, Coulomb, electron exchange and ion correlation have to be considered \cite{salpeter}. Also, the electron-capture (inverse beta-decay) has been shown to alter the known EoS and limits on the compact stars \cite{hamada} investigations of which is out of the scope of the current study.

\section{Conclusion and Summary}\label{conclusion}

Using the quantum magnetohydrodynamics and magnetohydrostatic models we showed that in a relativistically degenerate magnetized stellar objects such as white dwarfs the magneto-gravitational collapse is possible. The collapse was shown to be consistent with Landau orbital ferromagnetism with magnetic equation of state $B=\beta \rho^{2s/3}$ on some conditions. It was also revealed that, a critical $\beta_{cr}=\sqrt{2\pi}$ exists above which the plasma is hydrostatically unstable which comparing with the previous results indicates that in the LOFER plasma state quantum collapse and magnetogravity collapse can not take place at the same time. Furthermore, in a stable LOFER plasma it was shown that the Chandrasekhar mass-limit may critically depend on the value of the $\beta$ and $s$-parameters. These findings reveal the inevitable role of magnetism on the evolution of dense stellar objects.

\section{Acknowledgment}
Author is grateful to Prof. D. Garfinkle for providing the code for calculation of the mass-limit and constructive comments.

\newpage

\textbf{FIGURE CAPTIONS}

\bigskip

Figure-1

\bigskip

The surface in $\beta$-$x$-$s$ space showing where the quantum collapse is possible.

\bigskip

Figure-2

\bigskip

Variation of the value of $\eta_0$ with respect to change in the central density parameter $x_c$ for unmagnetized ($\beta=0$) case. The thickness of the curves is used as a measure for the value of the varied parameter.

\bigskip

Figure-3

\bigskip

The famous Chandrasekhar mass-radius curve for the case of unmagnetized ($\beta=0$) plasma indicating the Chandrasekhar mass-limit of $M_{Ch}\simeq1.43M_{S}$.

\bigskip

Figure-4

\bigskip

Variation of the value of $\eta_0$ with respect to change in the central density parameter $x_c$ for magnetized case ($B=\beta\rho^{2s/3}$) with ($s=2$) showing magnetohydrostatic instability of this model for $x_c>1$. The thickness of the curves is used as a measure for the value of the varied parameter.

\bigskip

Figure-5

\bigskip

Variation of the value of $\eta_0$ with respect to change in the central density parameter $x_c$ for magnetized case ($B=\beta\rho^{2s/3}$) with ($s=-1$) showing unphysical results for this model. The thickness of the curves is used as a measure for the value of the varied parameter.

\bigskip

Figure-6

\bigskip

Variation of the value of $\eta_0$ with respect to change in the central density parameter $x_c$ for magnetized case ($B=\beta\rho^{2s/3}$) with ($s=1/2$) showing the magnetohydrostatic stability for the range of central plasma density parameter, $x_c\neq 0$. The thickness of the curves is used as a measure for the value of the varied parameter.

\bigskip

Figure-7

\bigskip

Variation of the value of $\eta_0$ for fixed large central density parameter $x_c$ and varied magnetic parameter, $\beta^2<2\pi$, for magnetized case ($B=\beta\rho^{2s/3}$) with ($s=1/2$). The plot indicates that in this model the Chandrasekhar mass-limit does not depends on the value of the magnetic parameter, $\beta$. The thickness of the curves is used as a measure for the value of the varied parameter.

\bigskip

Figure-8

\bigskip

Variation of the value of $\eta_0$ for fixed large central density parameter $x_c$ and varied magnetic parameter, $\beta^2<2\pi$, for magnetized case ($B=\beta\rho^{2s/3}$) with ($s=1$) showing the magnetohydrostatic stability for the whole range of central plasma density parameter, $x_c$. The plot also shows that in this model the Chandrasekhar mass-limit depends strongly on the value of the magnetic parameter, $\beta$. The thickness of the curves is used as a measure for the value of the varied parameter.

\bigskip

Figure-9

\bigskip

Variation of the value of the Chandrasekhar mass-limit with the value of the ferromagnetic parameter, $\beta$ for the case of $s=1$. The horizontal line indicates ordinary mass-limit $M/M_S\simeq1.43217$.

\bigskip

Figure-10

\bigskip

The Chandrasekhar mass-radius curves for ordinary LOFER ($s=1$) with $\beta/\sqrt{2\pi}=0$ (rectangles), $\beta/\sqrt{2\pi}=0.1$ (empty circles), and $\beta/\sqrt{2\pi}=0.5$ (filled circles) for the case $s=1$.

\bigskip

\end{document}